\documentclass[vecphys]{svmult}

\usepackage{makeidx}         
\usepackage{graphicx}        
\usepackage{multicol}        
\usepackage[bottom]{footmisc}

\makeindex             

\begin{document}

\title*{$B \to K^* \ell^+ \ell^-$ as a probe of Universal Extra Dimensions}
\author{Rossella Ferrandes}

\institute{Phys. Dept. University of Bari and INFN, via Orabona 4,
I-70126 Bari, Italy \texttt{rossella.ferrandes@ba.infn.it}}
\maketitle

\section{Introduction}
\label{sec:1}

 The idea of the existence of \textit{extra dimensions} has recently obtained a
lot of attention \cite{rev}. In part, this interest is because the
scale at which the extra-dimensional effects can be relevant could
be around a few TeV, even hundreds of GeV in some cases, clearly a
challenging possibility for the next generation of accelerators.
Moreover, this new point of view has permitted to study many
long-standing problems (as the hierarchy problem) from a new
perspective.

\noindent An interesting model is that proposed by Appelquist,
Cheng and Dobrescu with so-called universal extra dimensions (UED)
\cite{ACD}, which means that all the Standard Model (SM) fields
may propagate in one or more compact extra dimensions. The
compactification of the extra dimensions introduces in the
four-dimensional description of the theory an infinite tower of
states for every field. Such states are called Kaluza-Klein (KK)
particles and their masses are related to compactification radius
according to the relation $ m_n^2=m_0^2+\frac{n^2}{R^2} $, with
$n=1, 2,...$. We consider the simplest Appelquist, Cheng and
Dobrescu (ACD) scenario, characterized by a single extra
dimension. It presents the remarkable feature of having only one
new parameter with respect to SM, the radius $R$ of the
compactified extra dimension.

\noindent Rare $B$ transitions can be used to constrain this
scenario \cite{Agashe:2001xt}. Buras and collaborators have
investigated  the impact of universal extra dimensions  on the
$B^0_{d,s}-\bar B^0_{d,s}$ mixing mass differences,  on the CKM
unitarity triangle  and on inclusive $b \to s$ decays for which
they have computed the
 effective Hamiltonian  \cite{Buras:2002ej,Buras:2003mk}. In particular, it was found that ${\cal B}(B \to X_s \gamma)$
allowed to constrain ${1/ R} \ge 250$ GeV, a bound updated by  a
more recent analysis to $1/R \ge 600$ GeV at 95$\%$ CL, or to $1/R
\ge 330 $ GeV at 99$\%$ CL \cite{Haisch:2007vb}.

\noindent In \cite{noi} several $B_{d,s}$ and $\Lambda_b$ decays
induced by $b \to s$ transitions were analyzed, finding that in
many cases the hadronic uncertainties do not hide the dependence
of the observables on $R$. In the following Sections we shall
discuss some of these results, such as the dependence on $R$ of
the branching ratio, the forward-backward asymmetry and the $K^*$
helicity distributions for the decay modes $B \to K^* \ell^+
\ell^-$, with $\ell^-=e^-,\mu^-$, and the tau polarization
asymmetries for the mode $B \to K^* \tau^+ \tau^-$.

\section{The decays $ B \to K^* \ell^+ \ell^-$}

In the Standard Model the effective $ \Delta B =-1$, $\Delta S =
1$ Hamiltonian governing  the rare transition $b \to s \ell^+
\ell^-$ can be written in terms of a set of local operators:
\begin{equation}
H_W\,=\,4\,{G_F \over \sqrt{2}} V_{tb} V_{ts}^* \sum_{i=1}^{10}
c_i(\mu) O_i(\mu) \label{hamil}
\end{equation}
\noindent where $G_F$ is the Fermi constant and $V_{ij}$ are
elements of the Cabibbo-Kobayashi-Maskawa mixing matrix; we
neglect terms proportional to $V_{ub} V_{us}^*$ since the ratio
$\displaystyle \left |{V_{ub}V_{us}^* \over V_{tb}V_{ts}^*}\right
|$ is of the order $10^{-2}$. We show only the operators $O_i$
which are relevant for the decays we consider here:
\begin{eqnarray}
O_7&=&{e \over 16 \pi^2} m_b ({\bar s}_{L \alpha} \sigma^{\mu \nu}
     b_{R \alpha}) F_{\mu \nu} \hspace{0.1cm},  \hspace{0.8cm}
O_9={e^2 \over 16 \pi^2}  ({\bar s}_{L \alpha} \gamma^\mu
     b_{L \alpha}) \; {\bar \ell} \gamma_\mu \ell \hspace{0.1cm}, \nonumber \\
O_{10}&=&{e^2 \over 16 \pi^2}  ({\bar s}_{L \alpha} \gamma^\mu
     b_{L \alpha}) \; {\bar \ell} \gamma_\mu \gamma_5 \ell \hspace{0.1cm},
\label{eff}
\end{eqnarray}
\noindent where $\alpha$, $\beta$ are colour indices,
$\displaystyle b_{R,L}={1 \pm \gamma_5 \over 2}b$, and
$\displaystyle \sigma^{\mu \nu}={i \over
2}[\gamma^\mu,\gamma^\nu]$; $e$ is the electromagnetic coupling
constant,while $F_{\mu \nu}$ denotes the electromagnetic field
strength tensor.

\noindent The Wilson  coefficients $c_i$ appearing in
(\ref{hamil}) are modified in the ACD model because the KK states
can contribute as intermediate states in penguin and box diagrams.
As a consequence, the Wilson coefficients can be expressed in
terms of functions $F(x_t,1/R)$, $x_t=\displaystyle{m_t^2 \over
M_W^2}$, which generalize the corresponding SM functions
$F_0(x_t)$ according to
$F(x_t,1/R)=F_0(x_t)+\sum_{n=1}^{\infty}F_n(x_t,x_n) $, where
$x_n=\displaystyle{m_n^2 \over M_W^2}$ and $m_n=\displaystyle{n
\over R}$.

\noindent The description of the decay modes $B \to K^* \ell^+
\ell^-$ involves the hadronic matrix elements of the operators
appearing in the effective Hamiltonians (\ref{hamil}). We use for
them two sets of results:  the first one,  denoted as set A,
obtained by  three-point QCD sum rules based on the short-distance
expansion \cite{Colangelo:1995jv};  the second one, denoted as set
B,   obtained  by  QCD sum rules based on the light-cone expansion
\cite{Ball:2004rg}.

\begin{figure}[h]
\begin{center}
 \includegraphics[width=0.35\textwidth] {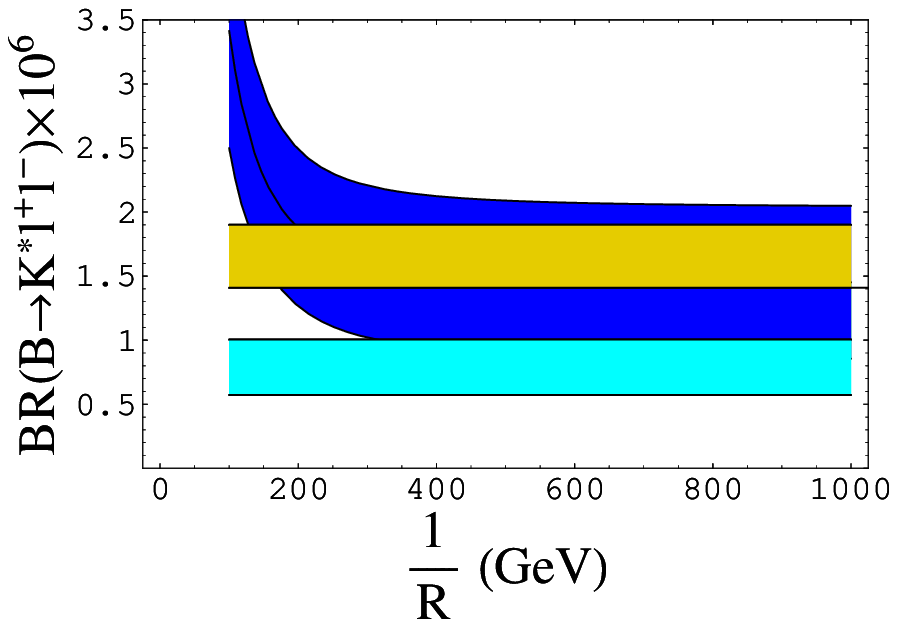}
 \hspace{0.5cm}
\includegraphics[width=0.35\textwidth] {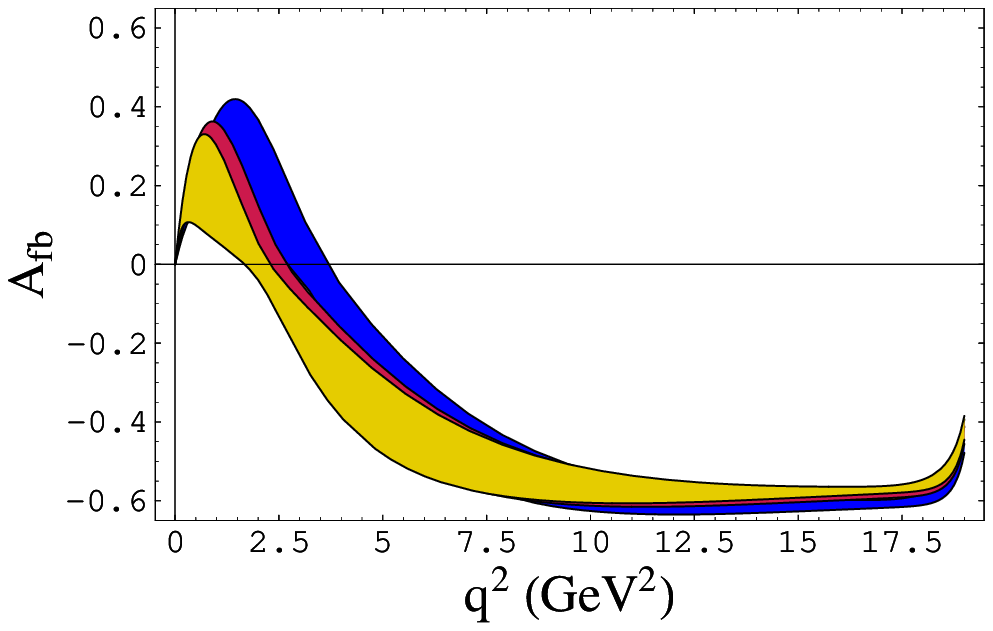}
\end{center}
\caption{\baselineskip=15pt Left: $BR(B \to K^* \ell^+ \ell^-)$
versus $\displaystyle{1 \over R}$ using set B of form factors. The
two horizontal regions correspond to  BaBar \cite{Aubert:2005cf}
(lower band) and Belle (upper band) \cite{Abe:2004ir} results.
Right: forward-backward lepton asymmetry in $B \to K^* \ell^+
\ell^-$  versus $\displaystyle{1 \over R}$ using set A. The dark
band correspond to the SM results, the intermediate band to
$1/R=250$ GeV, the light one to
 $1/R=200$ GeV. } \vspace*{1.0cm} \label{brkstar}
\end{figure}

\noindent With these ingredients we can calculate the branching
fraction as a function of $1/R$, as depicted in Fig.
\ref{brkstar}. The hadronic uncertainty is evaluated considering
the two set of form factors and taking into account their errors.
Comparing the theoretical prediction with the horizontal band
representing experimental data, we obtain that set A of form
factors does not allow to establish a lower bound on $1/R$, while,
as we can see in Fig. \ref{brkstar}, for set B one gets $1/R>200$
GeV. The present discrepancy between BaBar and Belle measurements
does not permit stronger statements.

\noindent Important information could be gained from the
 forward-backward
asymmetry, defined as \begin{equation} A^{FB} (q^2)=\displaystyle
{\displaystyle \int_0^1{d^2 \Gamma \over dq^2 d
cos\theta_\ell}dcos\theta_\ell -\int^0_{-1}{d^2 \Gamma \over dq^2
d cos\theta_\ell}dcos\theta_\ell \over\displaystyle \int_0^1{d^2
\Gamma \over dq^2 d cos\theta_\ell}dcos\theta_\ell
+\int^0_{-1}{d^2 \Gamma \over dq^2 d
cos\theta_\ell}dcos\theta_\ell} \; ,
 \label{asim_def}
\end{equation}
where $\theta_\ell$ is the angle between the $\ell^+$ direction
and the $B$ direction in the rest frame of the lepton pair (we
consider  massless leptons). This asymmetry is a powerful tool to
distinguish between  SM and several extensions of it. Belle
Collaboration has recently provided the first measurement of such
an observable \cite{Abe:2005km}. We show in the right part of Fig.
\ref{brkstar} our
 predictions for the SM,  $1/R=250$ GeV  and
$ 1/R=200$ GeV. A relevant aspect is that the zero of ${\cal
A}_{fb}$ is sensitive
 to the compactification parameter, so that  its experimental determination  would  constrain $1/R$.

\begin{figure}[htb]
\begin{center}
\includegraphics[width=0.35\textwidth] {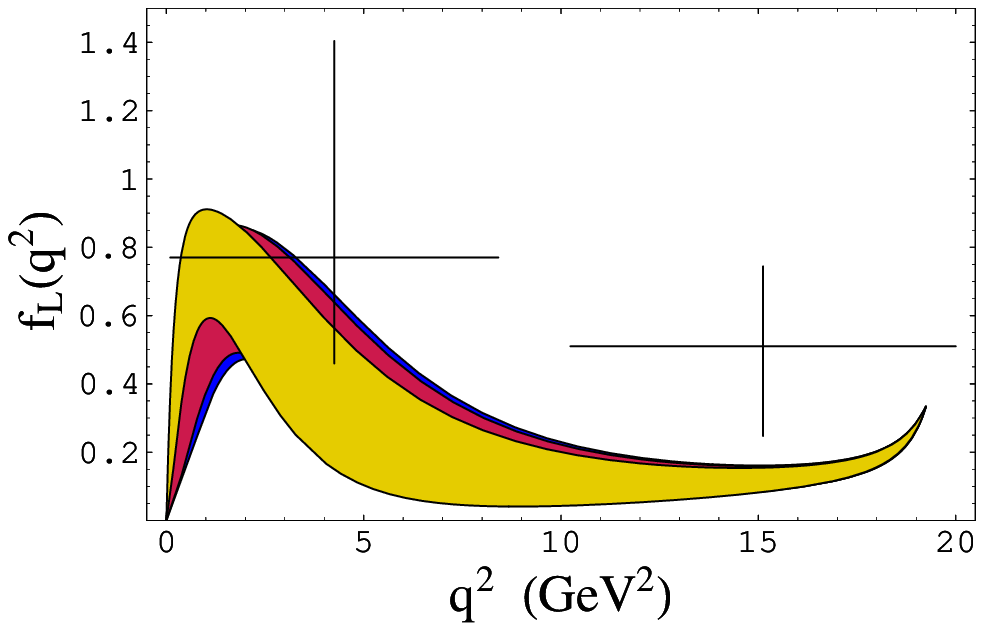}
\hspace{0.5cm}
\includegraphics[width=0.35\textwidth] {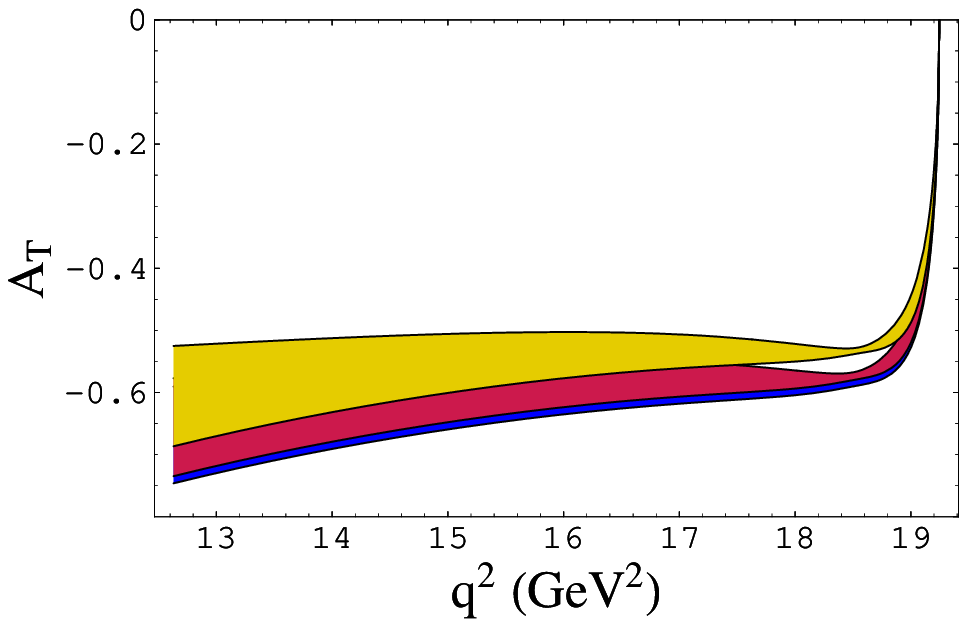}
\end{center}
\caption{\baselineskip=15pt Left: longitudinal $K^*$ helicity
fraction in $B \to K^* \ell^+ \ell^-$ obtained using  set A of
form factors.
 Right: Transverse
$\tau^-$ polarization asymmetry in $B \to K^* \tau^+ \tau^-$
 for set A of form factors. The  dark region is obtained in  SM;
 the intermediate one for $1/R=500$ GeV, the light one for $1/R=200$ GeV. } \vspace*{1.0cm}
\label{fui}
\end{figure}

\noindent We  investigate another  observable,
 the fraction of longitudinal $K^*$ polarization in  $B \to K^{*} \ell^+ \ell^-$, for which
 a new measurement in two bins of momentum transfer to the lepton pair is available  in case
 of $\ell=\mu, e$ \cite{Aubert:2006vb}:
\begin{eqnarray} f_L&=&0.77^{+0.63}_{-0.30}\pm 0.07 \hskip 1cm 0.1
\leq q^2 \leq 8.41 \,\,\, GeV^2 \,\,\,\,  \nonumber\\
f_L&=&0.51^{+0.22}_{-0.25}\pm 0.08 \hskip 1cm  q^2 \geq 10.24
\,\,\, GeV^2 \label{flexp} \,\,\, . \end{eqnarray}
    The dependence of  this quantity on the compactification parameter provides us with
  another possibility to constrain  the universal extra dimension
  scenario. In fact, we obtain that the value $q^2$ where this
  distribution has a maximum is sensitive to $R$, as we can see in
  the left part of  Fig. \ref{fui}.

\section{Lepton polarization asymmetries in $B \to K^* \tau^+ \tau^-$}
As first noticed in \cite{Hewett:1995dk},  the process $B \to K^*
\tau^+ \tau^-$ is of great interest due  to the possibility of
measuring lepton polarization asymmetries which  are sensitive to
the structure of the interactions, so that they
 can be used to test  the Standard Model and its extensions.

\noindent To compute lepton polarization asymmetries  for $B$
decays in $\tau$ leptons  we consider the spin vector $s$ of
$\tau^-$, with $s^2=-1$ and $k_1 \cdot s=0$,   $k_1$ being the
$\tau^-$ momentum. In the rest frame of the $\tau^-$ lepton three
orthogonal unit vectors: $e_L$, $e_N$ and $e_T$  can be defined,
corresponding to the longitudinal $s_L$, normal $s_N$ and
transverse $s_T$ polarization vectors:
\begin{eqnarray} s_L &=& (0,e_L)=\left( 0, {\vec k_1 \over |\vec k_1|}
\right), \hspace{0.8cm}  s_N = (0,e_N)=\left( 0, {\vec p^\prime
\times \vec k_1 \over |\vec p^\prime \times \vec k_1|}
\right) , \nonumber \\
s_T &=& (0,e_T)=(0,e_N \times e_L) \,\,\,  . \label{spinsrf}
\end{eqnarray} \noindent In eq.(\ref{spinsrf})
$\vec p^\prime$ and $\vec k_1$ are respectively the $ K^* $ meson
and the $\tau^-$ three-momenta in the rest frame of the lepton
pair.  Choosing  the $z$-axis directed as the $\tau^-$ momentum in
the rest frame of the lepton pair: $k_1=(E_1,0,0,|\vec k_1|)$ and
boosting the spin vectors $s$ in (\ref{spinsrf}) in the same
frame,  the normal and transverse polarization vectors $s_N,s_T$
remain unchanged: $s_N=(0,1,0,0)$ and $s_T=(0,0,-1,0)$, while the
longitudinal polarization vector becomes: $ s_L={1 \over
m_\tau}(|\vec k_1|,0,0,E_1) \,\,\, . \label{sl} $ For each value
of the squared momentum transfered to the lepton pair, $q^2$, the
polarization asymmetry for the negatively charged $\tau^-$ lepton
is defined as: \begin{equation} {\cal A}_A(q^2)=\displaystyle{ {d
\Gamma \over dq^2}(s_A)-{d \Gamma \over dq^2}(-s_A) \over {d
\Gamma \over dq^2}(s_A)+{d \Gamma \over dq^2}(-s_A) }
\label{def-pol}
\end{equation} with $A=L,T$ and $N$. In the right part of Fig.
\ref{fui} the transverse polarization asymmetry ${\cal A}_T$ is
shown for different values of $R$. It decreases (in absolute
value) by nearly $15\%$ with the decrease of $1/R$ down to
$1/R=200$ GeV.

\noindent In deriving the expressions of polarization asymmetries
it is possible to exploit some relations among form factors that
can be obtained in the large energy limit of the final meson  for
$B$ meson decays to a light hadron \cite{scet}. We obtain that, as
a consequence of such relations, the polarization asymmetries
become independent of form factors; this is a remarkable
observation, which renders the polarization asymmetries important
quantities to measure.

\section{Conclusions}
We have analyzed the  branching
 fraction as well as the forward-backward lepton asymmetry
 in $B \to K^* \ell^+  \ell^-$, founding that these observables are promising in order to constrain $1/R$. We have also considered the longitudinal $K^*$ helicity
fractions, for which some measurements are already available when
the leptons in the final state are $\ell=e,\mu$. For the mode $B
\to K^* \tau^+ \tau^-$, we have found that the dependence of the
$\tau^-$ polarization asymmetries
 on $\displaystyle 1/R$ is mild but still observable, the most
sensitive ones being the transverse asymmetry. Finally, during our
investigation we have shown that in the exclusive modes the
polarization asymmetries are free of hadronic uncertainties if one
considers the Large Energy limit for the light hadron in the final
state.



\end{document}